\documentclass[12pt,letterpaper,copyedit]{article}
\usepackage{osajnl2}
\usepackage{psfrag}
  
\definecolor{vert}{cmyk}{0.7,0,.7,0.5}
\definecolor{rouge}{cmyk}{0,1,1,.3}

\begin{document}
\title{Generation of correlated photons in hydrogenated amorphous silicon waveguides}
\author{S. Clemmen$^{1}$, A. Perret$^{2}$, S. K. Selvaraja$^3$,  W. Bogaerts$^3$, D. van Thourhout$^3$, R. Baets$^3$, Ph. Emplit$^4$ and S. Massar$^1$}
\address{$^1$ Laboratoire d'Information Quantique, CP 225, Universit\'{e} Libre de Bruxelles (U.L.B.), Boulevard du Triomphe, B-1050 Bruxelles, Belgium}
\address{$^2$ D\'epartement de Physique, \'Ecole normale sup\'erieure, 24 rue Lhomond, 75005 Paris, France}
\address{$^3$ Photonics Research Group, INTEC-department, Ghent University - IMEC,  Sint-Pietersnieuwstraat 41, 9000 Gent, Belgium}
\address{$^4$ Service OPERA-Photonique, CP 194/5, Universit\'e Libre de Bruxelles (U.L.B.), avenue F.D. Roosevelt 50,  1050 Brussels, Belgium}
\vspace{0.1cm}
\begin{center}\footnotesize{\textit{* Corresponding author sclemmen@ulb.ac.be}}\end{center}
\begin{abstract}
We report the first observation of correlated photons emission in hydrogenated amorphous silicon waveguides. We compare this to photon generation in crystalline silicon waveguides with same geometry. In particular we show that amorphous silicon has a higher nonlinearity and competes with crystaline silicon despite of higher loss. 
\hspace{-0.3cm} 
\end{abstract}
\ocis{(270.0270) Quantum optics; (190.4390) Nonlinear optics, integrated optics; (190.4380) Nonlinear optics, four-wave mixing}
%
\section{Introduction}
The traditional way of producing photon pairs is based on parametric downconversion in optical materials with a $\chi^{(2)}$ nonlinearity. Nevertheless, over the past decade there has been increasing interest in photon pair generation based on four-wave mixing (FWM). The latter types of sources were initially demonstrated in standard optical fibers~\cite{fiberpairs1} and microstructured fibers~\cite{fiberpairs5}. However, except when special precautions are taken, such as cooling the fibers, these fiber sources suffer from high noise level due to Raman scattering. More recently, photon pair sources based on FWM in crystalline silicon (c-Si) nanophotonic waveguides have been reported~\cite{kumar,ntt,takesue2,harada,clemmen}. {\em A priori}, in this case, one would expect c-Si souce of pairs to be noise-free, as there is negligible Raman gain in bulk silicon at the wavelength at which photon pairs are generated. However, a carefull study~\cite{clemmen,soi_noise} shows that uncorrelated photons do constitute a source of noise in this case, but almost two orders of magnitude smaller than in the case of fibers.
In parallel with the development of c-Si nanophotonics, the past years have seen the emergence of hydrogenated amorphous-silicon (a-Si:H) nanophotonics. a-Si:H has a number of potential advantages with respect to c-Si, which justifies the current interest in this technology. First of all, the material properties of deposited a-Si:H can be tuned by adjusting the deposition parameters; this could affect Raman scattering as well as the bandgap energy. These flexibilities make a-Si:H an attractive platform for nonlinear applications. Secondly, as a-Si:H can be deposited at relatively low temperatures, it could be deposited on many different substrates while keeping compatibility with the CMOS process. Furthermore, a-Si:H waveguides can be stacked into 3-dimensional optical circuits, whereas c-Si waveguides are restricted to planar architectures. 
Finally, a-Si:H waveguides can now be manufactured with losses comparable to c-Si waveguides~\cite{selvaraja}. As a-Si:H waveguides, have typically the same size as c-Si waveguides, and as a-Si:H material has  $\chi^{(3)}$ nonlinearity comparable to c-Si~\cite{asi-ultrafast, asi-optical}, the efficiency of the photon pair generation process should be comparable in both structures, and both c-Si and a-Si:H waveguides appear to be potentially attractive sources of photon pairs. 
%
In the present work we report the first observation of photon pair generation in a-Si:H nanophotonic waveguides at telecommunication wavelengths. In particular, we carry out a comparison between the rate of of photon pair generation in c-Si and a-Si:H waveguides. 
\section{Sample preparation and Experimental setup}
\begin{figure}
\begin{center}
\psfrag{laser}[cc][cc][1][0]{\footnotesize{laser}}
\psfrag{edfa}[cc][cc][1][0]{\footnotesize{edfa}}
\psfrag{p-c}[cc][cc][1][0]{\footnotesize{pc}}
\psfrag{dmux}[cc][cc][1][0]{\footnotesize{dmux1}}
\psfrag{d-mux}[cc][cc][1][0]{\footnotesize{dmux2}}
\psfrag{power-m}[cc][cc][1][0]{\footnotesize{P-meter}}
\psfrag{sss}[cc][cc][1][0]{\footnotesize{Si-wire}}
\psfrag{f-f}[cc][cc][1][0]{\footnotesize{ff}}
\psfrag{col}[cc][cc][1][0]{\footnotesize{col}}
\psfrag{sspd}[cc][cc][1][0]{\footnotesize{sspd}}
\psfrag{TDC}[cc][cc][1][0]{\footnotesize{tdc}}
\psfrag{d-e}[cc][cc][1][0]{\footnotesize{delay}}
\psfrag{bpf}[cc][cc][1][0]{\footnotesize{bpf}}
\psfrag{bbf}[cc][cc][1][0]{\footnotesize{bbf}}
\psfrag{Atn}[cc][cc][1][0]{\footnotesize{Atn}}
\psfrag{K}[cc][cc][1][0]{\footnotesize{$\sharp$}}
\psfrag{d-e2}[cc][cc][1][0]{\footnotesize{time}}
\psfrag{histo}[cc][cc][1][0]{\footnotesize{histo}}
\includegraphics[width=8cm]{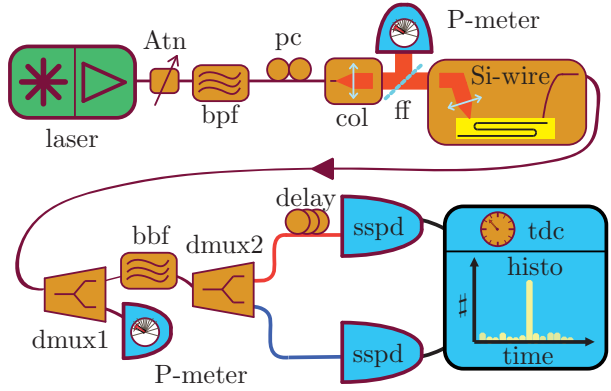}
\caption{(Color online)
Coincidences measurement : laser~: pumping CW-beam at 1539.8~nm amplified by an EDFA; Atn~: tunable attenuator; pc~: polarization controller; bpf~: bandpass filter centered at pump wavelength; ff~: switch miror,  col~: collimation package; P-meter~: power meter; dmux1~: demultiplexer add $\&$ drop filter for the pump band; bbf~: bandblock filter; dmux2~: Stokes/anti-Stokes selector; sspd~: superconductor single photon detectors; tdc~: time-to-digital converter provides a result which is a histogram.
}
\label{fig:setup}
\end{center}
\end{figure}

The silicon waveguides used in the present experiment were fabricated on a silicon wafer with 220~nm thick a-Si:H on top of 2000~nm of $\textrm{SiO}_2$. Amorphous silicon was deposited using low temperature ($300^{\circ}$C) plasma enhanced chemical vapor deposition process on top of 2~$\mu$m $\textrm{SiO}_2$ layer~\cite{selvaraja}. After the a-Si:H layer deposition phase, single mode 500~nm wide and 11.2~mm long waveguides were defined using 193~nm optical lithography and dry etching~\cite{refshankar2}. Identical grating couplers were defined for in- and out-light-coupling. On the other hand, waveguides made of crystaline silicon were fabricated with same section on top of a 2~$\mu$m layer of $\textrm{SiO}_2$. Measurements made for amorphous (resp. crystalline) silicon waveguides revealed in/out-coupling losses around $8\pm1$~dB (resp. $6\pm0.5$~dB) and propagation loss around $4.5\pm0.5$~dB (resp. $2.5\pm0.5$~dB).
\\
The experimental setup relies on a coincidence measurement which is depicted in Fig.~\ref{fig:setup}. 
Photon pairs are generated while pumping the waveguides with a CW-beam at telecom wavelength. The power of the pump beam can be adjusted with a tunable attenuator so that for 0~dB attenuation, the power before incoupling is 10~mW. A bandpass filter (BPF) ensures absolute darkness at Stokes and anti-Stokes frequencies. This BPF is made of fiber bragg gratings, circulators and commercial DWDM add \& drop filters (100~GHz on ITU grid and centered at 1539.8~nm). Overall, the BPF provides extinction greater than 150~dB outside of the pump band  1538.9-1540.6~nm (pump band). Last DWDM filter of the BP has a short pigtail (10~cm) to limit Raman scattering in the fiber. 
\\
Correlated pairs are exhibited by deterministically splitting the photon pairs and optically delaying one of the photons. A first demultiplexer separated the pump beam from the pairs. A second demultiplexer selects two spectral band~: Stokes from 1541.5 to 1558.5~nm and anti-Stokes band from 1523 to 1538.5~nm.
Photons are detected thanks to superconducting single photon detectors (from Scontel) cooled down to $1.8 \pm 0.1$~K; efficiencies of detectors are $6.1\pm0.1 \%$ and $5\pm1 \%$~\cite{footnote1} while dark counts are $80 \pm 20$~Hz and $25 \pm 12$~Hz. The time difference between both detections is measured with a time-to-digital converter (TDC - Agilent Acquiris system) so that the entire detection system can resolve coincidences with 80~ps resolution (fwhm of a coincidence peak). TDC system collects all events, including single detection events, and sends them to a computer. This limits the rate at which coincidences can be measured by the TDC system and requires thus a calibration of the system for obtaining an absolute rate of coincidences. Furthermore, the absolute flux at anti-Stokes frequency is measured  by replacing the Time-to-digital convertor by an auxiliary counter. 
Stokes and anti-Stokes bands are chosen so that the pair flux is spectrally flat over the selected bandwidth.  Indeed theory predicts that the pair flux is given by a $\textrm{sinc}$ function which is flat for low values of its argument :
\begin{equation}
\Phi=
\frac{1}{2\pi} \int_{\Delta \omega}   \left| \gamma P L \:  \textrm{sinc} \left[   L \left( \frac{(\beta_2 \omega^2)^2}{4}+\beta_2 \omega^2 \gamma P \right)^{1/2} \right]   \right| ^2  \:  \textrm{d} \omega 
\label{eq:flux}
\end{equation}
where $\gamma$ is the third order nonlinearity coefficient of the waveguide (around $200\textrm{ W}^{-1} \textrm{m}^{-1}$ for c-Si), $\beta_2$ is the group velocity dispersion parameter of the waveguide (around $-1\textrm{ ps}^2 \textrm{m}^{-1}$ for waveguide section),  $\Delta \omega$ is the bandwidth of the demultiplexer used to collect Stokes and anti-Stokes photons, $P$ is the pump power in the waveguide, and $L$ is waveguide's length.
%
\section{Results}
\begin{figure}
\begin{center}
\psfrag{coinflux}[bc][tc][1][0]{\footnotesize{Pair flux ($10^2 \times$ Hz)}}
\psfrag{pumppower}[tc][bc][1][0]{\footnotesize{Pump power before incoupling (mW)}}
\psfrag{0}[cc][cc][1][0]{\footnotesize{0}}
\psfrag{2}[cc][cc][1][0]{\footnotesize{2}}
\psfrag{4}[cc][cc][1][0]{\footnotesize{4}}
\psfrag{6}[cc][cc][1][0]{\footnotesize{6}}
\psfrag{8}[cc][cc][1][0]{\footnotesize{8}}
\psfrag{10}[cc][cc][1][0]{\footnotesize{10}}
\psfrag{Y1}[cr][cr][1][0]{\footnotesize{0}}
\psfrag{Y2}[cr][cr][1][0]{\footnotesize{1}}
\psfrag{Y3}[cr][cr][1][0]{\footnotesize{2}}
\psfrag{Y4}[cr][cr][1][0]{\footnotesize{3}}
\psfrag{Y5}[cr][cr][1][0]{\footnotesize{4}}
\psfrag{Y6}[cr][cr][1][0]{\footnotesize{5}}
\includegraphics[width=8cm]{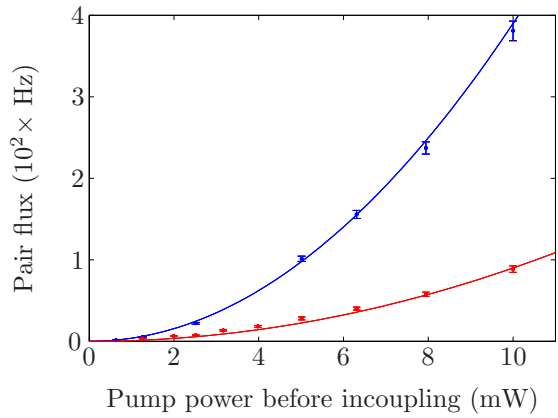}
\caption{(Color online) Detected coincidence rate versus pump power before incoupling in the c-Si waveguide (blue, top curve)  and in the a-Si:H waveguide (red, bottom curve).  Error bars come from  Poisson statistics. Regression curves are quadratic fits following Eq.~\ref{eq:flux}.
}
\label{fig:pairflux}
\end{center}\end{figure}
\begin{figure}
\begin{center}
\psfrag{CAR}[bc][tc][1][0]{\footnotesize{CAR}}
\psfrag{coinflux}[tc][bc][1][0]{\footnotesize{Measured coincidences rate (Hz)}}
\psfrag{x0}[cc][cc][1][0]{\footnotesize{$1$}}
\psfrag{x1}[cc][cc][1][0]{\footnotesize{$10$}}
\psfrag{x2}[cc][cc][1][0]{\footnotesize{$10^2$}}
\psfrag{x3}[cc][cc][1][0]{\footnotesize{$10^3$}}
\psfrag{x4}[cc][cc][1][0]{\footnotesize{$10^4$}}
\psfrag{Y1}[cr][cr][1][0]{\footnotesize{$1$}}
\psfrag{Y2}[cr][cr][1][0]{\footnotesize{$10$}}
\psfrag{Y3}[cr][cr][1][0]{\footnotesize{$10^2$}} 
\includegraphics[width=8cm]{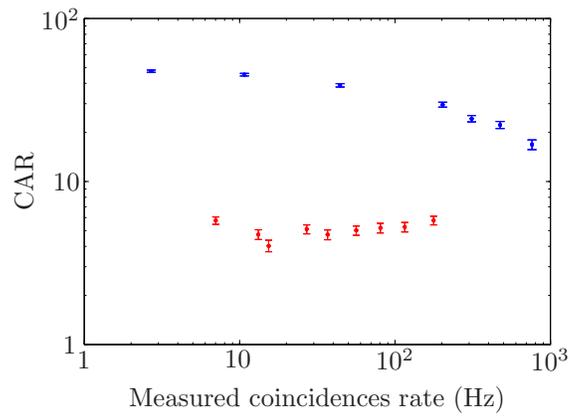}
\caption{(Color online) Coincidences-to-accidental detections ratio (CAR) versus the measured pair flux in the a-Si:H waveguide (red, bottom curve) and in the c-Si waveguide (blue, top curve). A-si:H exhibits a reduction by almost one order of magnitude of CAR. Error bars come from strandard error on Poisson statistic. 
}
\label{fig:car}
\end{center}\end{figure}

Comparison of pair fluxes generated in a-Si:H and c-Si in presented in Fig.~\ref{fig:pairflux}. As expected, the pair flux $\Phi$ grows quadratically with pump power $P$. We also expected that the pair flux generated in a-Si:H silicon is lower than in c-Si because of higher propagation and in-coupling loss.   
As in-coupling loss is around 2~dB higher, we would expect photon pair generation efficiency to be reduced by $60\%$ in comparison to efficiency in c-Si. Furthermore, propagation loss and out-coupling loss are both 2~dB higher too, that would result in another reduction by $84\%$ (-8~dB)  of the detected coincidences. Overall, those additional losses should reduce detection rate by $95\%$ for a given input power. Fig.~\ref{fig:pairflux} indicates the photon pair generation is around four times lower in the amorphous silicon waveguide in comparison to c-Si waveguide. This implies a  Kerr nonlinearity coefficient in a-Si:H higher by a factor 2.2. This is very close to result obtained by self phase modulation~\cite{bart} and also compatible with a recent study in similar waveguides~\cite{asi-optical}. Because of higher propagation loss, the figure of merit is certainly worse than in c-Si.  This is confirmed by Fig.~\ref{fig:car} that shows comparison of coincidences to accidental ratio (CAR) for c-Si and a-Si:H. We find a reduction of CAR by up to one order of magnitude. In Fig.~\ref{fig:car}, dark counts are so low that the main sources of accidental coincidences are either broken pairs or noise from the source itself.
\begin{figure}
\begin{center}
\psfrag{flux}[bc][tc][1][0]{\footnotesize{Detected flux (Hz)}}
\psfrag{pumppower}[tc][bc][1][0]{\footnotesize{Pump power in waveguide (mW)}}
\psfrag{x0}[cc][cc][1][0]{\footnotesize{$10^{-2}$}}
\psfrag{x1}[cc][cc][1][0]{\footnotesize{$10^{-1}$}}
\psfrag{x2}[cc][cc][1][0]{\footnotesize{1}}
\psfrag{Y1}[cr][cr][1][0]{\footnotesize{$10^3$}}
\psfrag{Y2}[cr][cr][1][0]{\footnotesize{$10^4$}}
\psfrag{Y3}[cr][cr][1][0]{\footnotesize{$10^5$}}
\psfrag{Y4}[cr][cr][1][0]{\footnotesize{$10^6$}}
\includegraphics[width=8cm]{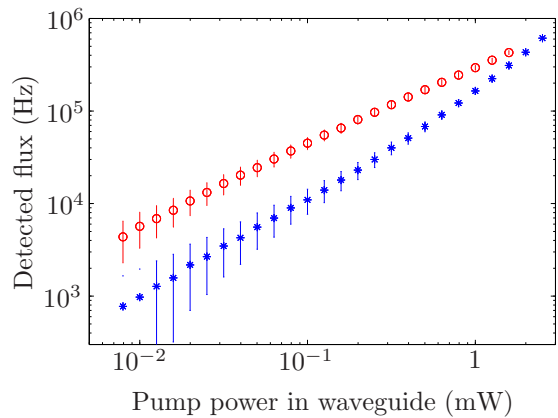}
\caption{(Color online) Measured photon flux versus input power in a-Si:H (red, top curve) and c-Si (blue, bottom curve) in the anti-Stokes band. The flux generated is slightly higher in a-Si:H despite higher propagation and coupling losses. Error bars come from strandard error on Poisson statistics.
}
\label{fig:flux}
\end{center}\end{figure}
Fig.~\ref{fig:flux} compares the fluxes generated in a-Si:H and c-Si in anti-Stokes band. The higher flux generated in a-Si:H waveguide despite higher propagation loss clearly indicates either a higher nonlinearity or a higher source of noise, or both.  Elsewhere~\cite{soi_noise} we study the origin of the weak noise that arises in c-Si and demonstrated that this noise is not due to carrier dynamics but is related to a thermal population of phonons, probably Raman scattering.  In a-Si:H, this noise source could be enhanced because of a broader Raman resonance as it was demonstrated in porous Si~\cite{ramanporous}.
Note that we are able to observe and quantify the noise in Fig.~\ref{fig:car} because we operate in a CW regime, collect Stokes and anti-Stokes photons in a wide spectral band ($\pm 15$~nm), and use detectors with very low dark count rates.
\section{Conclusion}
We have shown that amorphous silicon, as well as crystalline silicon, nanophotonics is an interesting plateforms for quantum optics as it provides efficient and low noise source of photon pairs. We have observed that amorphous silicon has a higher Kerr nonlinearity than c-Si but also suffer from higher loss which results in an overall efficiency which is not as good as in c-Si. Nevertheless, a-Si:H is more versatile, as it can be deposited on many substrates while keeping compatibility with CMOS process, and as it allows for 3-dimensional architectures. This would certainly eases integration with other components required for Linear Optics Quantum Computer (like detectors) while keeping mechanical stability and straightforward combination with linear components (interferometers, filters, beam splitters, ...). 
%
\section*{Acknowledgments}
We acknowledge the support of the \textit{Fonds pour la formation \`a la Recherche dans l'Industrie et dans l'Agriculture} (FRIA, Belgium), of the \textit{Interuniversity Attraction Poles Photonics@be Programme} (Belgian Science Policy) under grant IAP6-10. We thank Bart Kuyken for regular feedback about a-Si:H properties. Wim Bogaerts acknowledges the Flemish Research Foundation (FWO-Vlaanderen) for a postdoctoral fellowship.
%


\begin{thebibliography}{99}
\bibitem{footnote1} Higher uncertainty on second detector's efficiency is due to absence of polarization controller before the superconducting chip.
\end{thebibliography}
\end{document}